\documentclass[%
 reprint,
 amsmath,amssymb,
 aps,
prl,
]{revtex4-2}

\usepackage{graphicx}
\usepackage{dcolumn}
\usepackage{bm}
\usepackage[caption=false]{subfig}
\usepackage{hyperref}
\hypersetup{unicode=true, colorlinks=true, linkcolor=blue, citecolor=blue}

\begin{document}

\preprint{APS/123-QED}

\title{Quantum Transport and Molecular Sensing in Reduced Graphene Oxide Measured with Scanning Probe Microscopy}

\author{Julian Sutaria}

\author{Cristian Staii}%
 \email{cstaii01@tufts.edu}
\affiliation{Department of Physics and Astronomy, Tufts University, Medford, MA, 02155, USA
}%

\date{\today}

\begin{abstract}
We report combined scanning probe microscopy and electrical measurements to investigate local electronic transport in reduced graphene oxide (rGO) devices. We demonstrate that quantum transport in these materials can be significantly tuned by the electrostatic potential applied with a conducting atomic force microscope (AFM) tip. Scanning gate microscopy (SGM) reveals a clear p-type response in which local gating modulates the source--drain current, while scanning impedance microscopy (SIM) indicates corresponding shifts of the Fermi level under different gating conditions. The observed transport behavior arises from the combined effects of AFM tip–induced Fermi-level shifts and defect-mediated scattering. These results show that resonant scattering associated with impurities or structural defects plays a central role and highlight the strong influence of local electrostatic potentials on rGO conduction. Consistent with this electrostatic control, the device also exhibits chemical gating and sensing: during exposure to electron-withdrawing molecules (acetone), the source--drain current increases reversibly and returns to baseline upon purging with air. Repeated cycles over 15 minutes show reproducible amplitudes and recovery. Using a simple transport model we estimate an increase of about \(40\%\) in carrier density during exposure, consistent with p-type doping by electron-accepting analytes. These findings link nanoscale electrostatic control to macroscopic sensing performance, advancing the understanding of charge transport in rGO and underscoring its promise for nanoscale electronics, flexible chemical sensors, and tunable optoelectronic devices.
\end{abstract}

\maketitle


\section{I. Introduction}
Graphene and its chemically modified derivatives are widely studied for applications in nanoscale electronics \citep{Novoselov2004, Geim2007, Chen2008, Wang2010, Urade2023}, optics \citep{Abid2018, Chang2013, Mueller2010, Loh2010}, mechanical systems \citep{Lee2008, Hu2022, Xuan2018, Li2019}, biosensing \citep{Kim2013, Ohno2009, Mohanty2008, Faruque2021, Poo-arporn2019, Salimiyan2023}, and chemical sensing \citep{Schedin2007, Dan2009, Fowler2009, Staii2005_DNA, Lu2010, Lu2011, Deng2012, Dua2010}. 
The unique two-dimensional honeycomb lattice structure of graphene provides a large surface exposure and tunable electronic structure, rendering the electrical response highly sensitive to changes in surface conditions, particularly the adsorption and desorption of gas molecules \citep{Novoselov2004, Geim2007, Chen2008, Wang2010, Urade2023, Schedin2007, Dan2009, Fowler2009, Staii2005_DNA, Lu2010, Lu2011, Deng2012, Dua2010}.  Devices based on pristine or chemically functionalized graphene accordingly detect a wide range of gases at very low concentrations \citep{Schedin2007, Dan2009, Fowler2009, Staii2005_DNA, Lu2010, Lu2011, Deng2012, Dua2010}, underscoring the suitability of graphene-related materials for nanoscale chemical-vapor sensing. 
Within this materials family, reduced graphene oxide (rGO) platelets have emerged as a cost-effective and scalable platform for graphene-based electronic sensors \citep{Lu2011, Deng2012, Dua2010, Robinson2008, Lee2019, Oh2014, Cai2014, Mao2010, Hasegawa2014}. Reduced Graphene Oxide (rGO) stands out as a significant member of the graphene family, primarily because it is one of the few variants that can be synthesized on a large scale, reaching kilogram-level production \citep{Rowley2018, Lu2009, Zhu2010}. Typically, electrically conductive rGO platelets are produced by treating water-dispersed graphene oxide with reducing agents such as hydrazine, sodium borohydride, or ascorbic acid \citep{Zhu2010, Stankovich2007, Si2008}. As an electronically hybrid material, rGO can be precisely tuned from an insulating state to a semiconductor through reduction processes \citep{Gilje2007, Sehrawat2018, Jung2008}. The resulting rGO retains functional groups, including carboxyl, alcohol, and dangling oxygen species, within its hexagonal carbon lattice. These functional groups enable rGO to interact with a broad spectrum of chemical analytes, facilitating charge transfer processes in which molecules act as either electron donors or acceptors. This interaction leads to noticeable variations in the electrical conductivity of rGO-based devices \citep{Dua2010, Robinson2008, Sehrawat2018, Jung2008, Kehayias2013}.
The ability to chemically modify rGO devices enables precise tuning of their electrical conductivity, optical transparency, and sensitivity/selectivity for analyte detection. However, unlike individual graphene sheets, research on electronic transport in rGO has only recently gained attention. Several challenges must be overcome before rGO-based devices can be widely used in applications such as chemical and biological sensors, transparent conductors, transducers, and smart composite materials. Critical issues include the need for a deeper understanding of quantum transport, the spatial distribution of charge carriers, charge transfer mechanisms between analytes and rGO, and the impact of contact resistance between rGO and metal electrodes. Addressing these challenges is crucial for optimizing the design and performance of rGO-based devices.
In our previous work, we used a scanning probe microscopy (SPM) technique called Kelvin probe microscopy to quantitatively analyze chemical gating effects in rGO field-effect transistor devices \citep{Kehayias2013}. These devices exhibited highly selective and reversible responses to analytes, with fast response and recovery times (within tens of seconds). We quantified charge transfer due to chemical doping when exposed to electron-acceptor (acetone) and electron-donor (ammonia) analytes. This approach enabled high-resolution mapping of surface potential and local charge distribution, as well as the direct extraction of contact resistance between rGO and metallic electrodes.

Here we combine electronic transport with SPM experiments to measure variations in surface potential, Fermi energy, and charge carrier density in rGO devices patterned on SiO$_2$/Si substrates. We show that these devices exhibit transport currents that are highly dependent on the local gate voltage applied by an atomic force microscope tip. Combined scanning gate and scanning impedance microscopy measurements indicate that this behavior arises from resonant electron scattering as the gate voltage shifts the Fermi level. In the same devices we also observe reversible chemical gating, where exposure to electron-withdrawing molecules increases the source to drain current and the current returns to baseline in air. We develop a theoretical model that quantifies the shift in the Fermi level and the change in carrier density between the peak and the resonant scattering states.

\section{II. Materials and Methods}

\textit{2.1 Sample Preparation}. RGO platelets were synthesized using a modified Hummers method, as previously reported \citep{Kehayias2013}. Graphite nanoplatelets were chemically treated to produce graphene oxide, which was then reduced using ascorbic acid (vitamin C). The resulting RGO powder was suspended in dimethyl formamide and further processed into a fine nanoplatelet suspension via sonication. To enable RGO assembly, Au electrodes were patterned on doped Si wafers with a 200 nm oxide layer using sputtering techniques. RGO-based devices were subsequently fabricated by assembling RGO platelets between the Au source and drain electrodes via dielectrophoresis \citep{Kehayias2013}. The distance between the
patterned Au electrodes (source-drain) is approximately 5 $\mu$m.

\begin{figure}
    \includegraphics[width=1\linewidth]{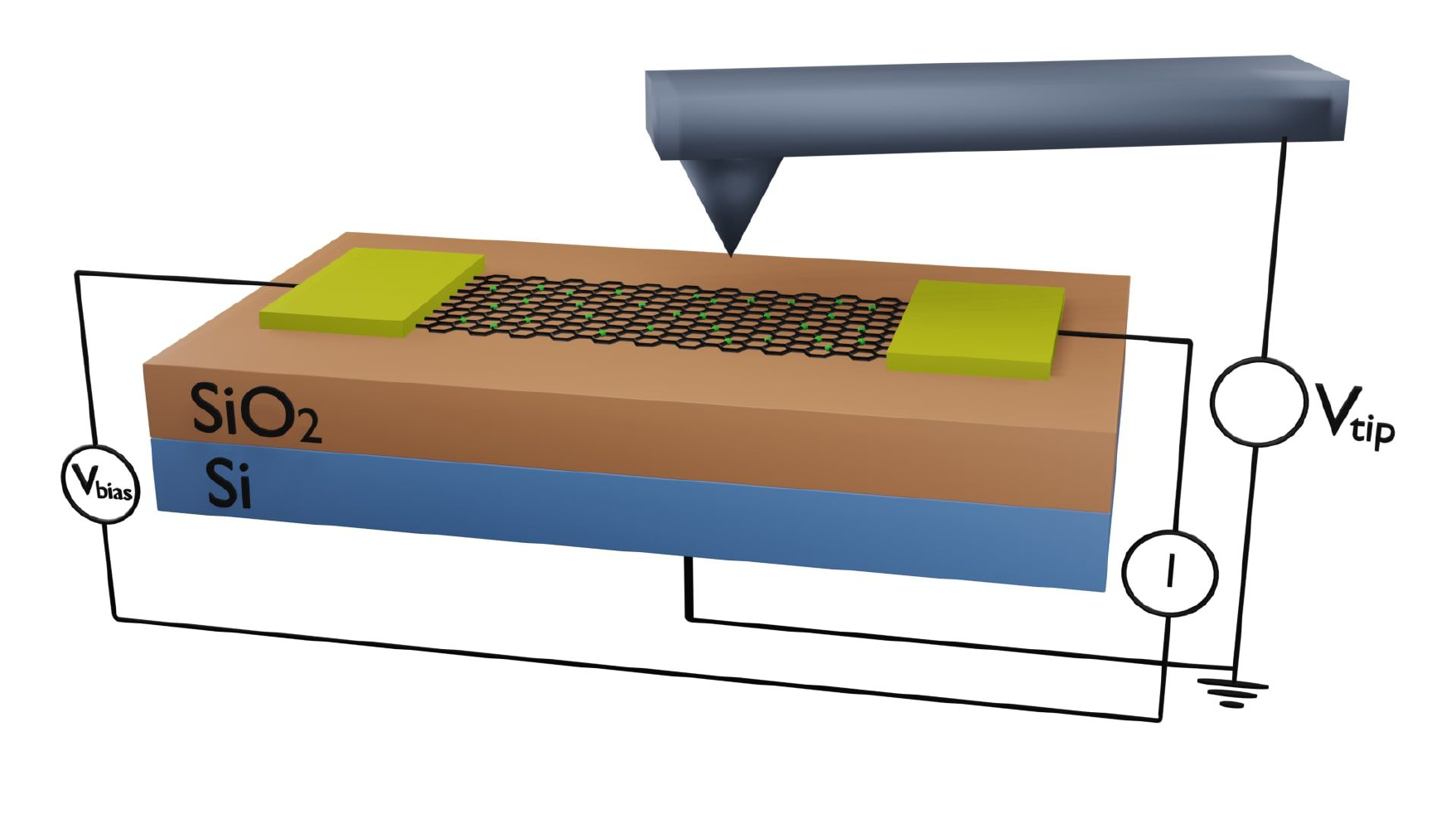}
    \caption{Schematic illustration of the Scanning Impedance Microscopy (SIM) and Scanning Gate Microscopy (SGM) experiments. A bias voltage is applied across the rGO sample between the source and drain Au electrodes. In SIM measurements, a voltage - biased AFM tip scans at a fixed height (50 nm) above the sample while its oscillation amplitude is recorded as a function of the tip position, providing a direct measurement of the local surface potential beneath the tip. In SGM measurements the transport current $I$ through the sample is measured as a function of the tip voltage. \label{fig1}}
\end{figure}

\textit{2.2 Scanning Impedance Microscopy and transport measurements}. Scanning Impedance Microscopy (SIM) and transport measurements are performed using the device schematics shown in Figure~\ref{fig1}. SIM is a dual-pass technique that can be used to measure the local surface potential of a sample \citep{Kehayias2013, Curtin2011, Kalinin2001, Yu2009}. In the first line scan, the tip acquires a topography profile in tapping mode. In the second line scan, the tip travels at a defined height above the surface.
During this second pass, both a DC voltage $V_{tip}$ is applied to the cantilever, and an AC bias voltage is applied to the sample:
\begin{equation}
V_{\mathrm{bias}} = V_{\mathrm{dc}} + V_{\mathrm{ac}} \sin(\omega t)
\label{F1}
\end{equation}
where $\omega$ is equal to the resonant frequency of the cantilever. This AC signal establishes a potential distribution within the sample, which subsequently generates a position-dependent electrostatic force on the tip, causing the cantilever to resonate. The first harmonic of the interaction force between the tip and the sample is proportional to the local surface potential beneath the tip $V(x,y)$: \citep{Kehayias2013,Curtin2011}:

\begin{equation}
F_{\omega} 
= \left(\frac{dC}{dz}\right)\bigl(V_{\mathrm{tip}} - V_{\mathrm{dc}}\bigr) \cdot V(x,y)
\ \cdot \sin(\omega t)
\label{Former2}
\end{equation}
where $C(z)$ is the tip--surface capacitance.

Thus the SIM image records the distribution of the surface potential along the sample surface by mapping the cantilever oscillation amplitude versus the tip position.

In addition, for the rGO surface we have that \citep{Curtin2011}:
\begin{equation}
e \,V(x,y) = W_{\mathrm{cn}} - E_{F},
\label{F3}
\end{equation}
where $e$ is the electron charge, $W_{\mathrm{cn}}$ is the (constant) work function of the charge-neutral rGO, and $E_{F}$ is the Fermi energy. Equations~(1)--(3) indicate that changes in $V(x,y)$ measured by SIM directly correspond to variations in the local Fermi energy $E_{F}$ of the rGO sample:

\begin{equation}
\Delta E_{F} \approx e\,\Delta V(x,y).
\label{F4}
\end{equation}

\textit{Capacitance ladder and quantum capacitance.} The tip--sample electrostatic coupling that underlies Equation~(2) involves both the geometric capacitance $C_{\mathrm{geo}}(z)$ (set by the tip geometry and scan height $h$) and the quantum capacitance $C_Q$ of the rGO sheet, $C_Q = e^2 D(E_F)$, where $D(E_F)$ is the electronic density of states (DOS) at the Fermi level \cite{Fang2007,Luryi1988}. For a graphene-like Dirac dispersion one has $C_Q/{A}= 2 e^2 |E_F|/(\pi (\hbar v_F)^2)$, with $v_F$ the band velocity for rGO \citep{Datta1995, Fang2007}. The tip--sample coupling thus forms a series combination with effective capacitance per unit area $1/C_{\mathrm{eff}} \;=\; 1/C_{\mathrm{geo}} + 1/C_Q$, which sets the lever arm $\eta \equiv C_{\mathrm{eff}}/C_Q \le 1$ that relates the applied tip potential to the Fermi--level shift in the rGO under the AFM tip. Using these expressions with representative values $|E_F|\sim 0.1$--$0.2$~eV and $v_F \simeq  10^5$~m/s \citep{Sehrawat2018, Jung2008, Kehayias2013}, we obtain $C_Q/A \sim 15$--$30~\mu\mathrm{F\,cm^{-2}}$, whereas a parallel--plate estimate for the geometric term at $h=50$ nm gives $C_{\mathrm{geo}}/A \approx \varepsilon_0/h \simeq 0.017~\mu\mathrm{F\,cm^{-2}}$. Thus $C_Q \gg C_{\mathrm{geo}}$ and $C_{\mathrm{eff}} \approx C_{\mathrm{geo}}$ in our geometry. Importantly, the measured SIM voltage $V(x,y)$ in Equation~(2) is the actual local surface potential. Consequently, combining Equations ~(3) and (4) remains valid. Quantum capacitance mainly enters when one converts an applied gate voltage to a carrier density change, in which case $\Delta n \approx (C_{\mathrm{eff}}/e)\,\Delta V_{\rm gate}$ (see the estimates of the change in carrier density below).

\textit{2.3 Scanning Gate Microscopy}. In Scanning Gate Microscopy (SGM) imaging mode, a conducting tip with an applied voltage ($V_{tip}$) is scanned at a fixed height above an electrically biased sample (with source-drain voltage $V_{bias}$), while the source-drain transport current $I$ is recorded as a function of the tip’s position (Figure~\ref{fig1}). Unlike a static backgate, which capacitively couples to the entire sample, the tip acts as a spatially localized gate with a controllable position. The resulting image, generated from the variation of the transport current $I$ with tip position, shows specific sample regions where the device exhibits a strong response to the tip-induced gating.  In addition to imaging mode, SGM can also operate in spectroscopy mode, where the tip remains fixed at a specific location while the transport current through the sample is measured as a function of tip voltage $I(V_{tip})$ or tip-sample distance $I(h)$. SGM  has been used to directly investigate conduction properties in nanoscale materials, including carbon nanotubes and graphene \citep{Staii2007, Staii2005_SGM, Freitag2002, Jalilian2011, Bockrath2001}.
For the measurements reported in this paper SIM and SGM measurements were taken on an Asylum Research MFP3D AFM using platinum – coated tips with curvature radius $R=20-30$ nm, quality factor $Q=150$ and spring constant $k=0.65-1$ N/m. The range for voltage applied to the tip was between -10V to 10V and the scan height was $h = 50$ nm.

\section{III. Results}

\begin{figure}
    \includegraphics[width=1\linewidth]{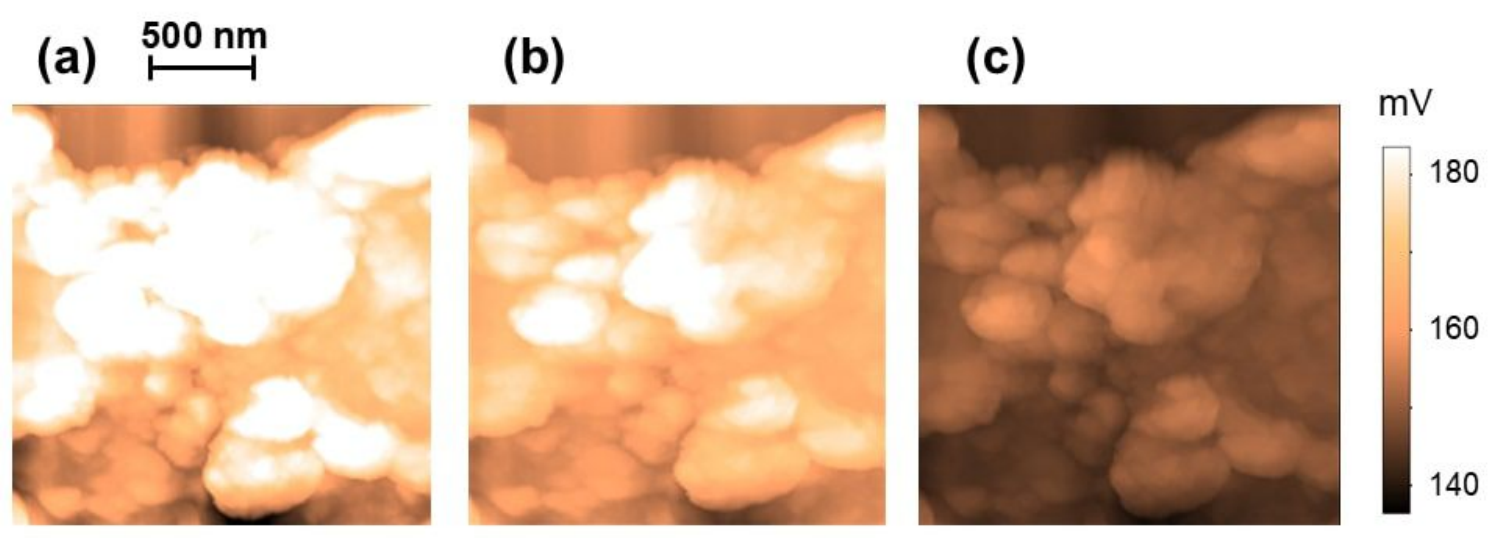}
    \caption{Scanning Impedance Microscopy images of the same region on an rGO sample. The images show the distribution of the surface potential corresponding to three different tip voltages: (\textbf{a}) $V_{tip}=-10V$ , (\textbf{b}) $V_{tip}=-2V$, and (\textbf{c}) $V_{tip}=+10V$ (\textbf{c}). The voltage scale bar on the right is the same for all three figures.  \label{fig2}}
\end{figure}

\textit{3.1 Scanning Impedance Microscopy Measurements.}
As discussed in the Materials and Methods section, SIM measures the distribution of the surface potential $V(x,y)$ along the sample surface with very high spatial resolution. Figure~\ref{fig2} shows examples of SIM images taken on the same rGO sample at three different tip voltages: $V_{tip}=-10V$ (Figure~\ref{fig2}a), $V_{tip}=-2V$ (Figure~\ref{fig2}b), and $V_{tip}=+10V$ (Figure~\ref{fig2}c). These measurements demonstrate that the voltage - biased tip simultaneously probes the local surface potential and changes the carrier density through the sample. Similar "tip-gating" effects have been reported on combined SIM - SGM experiments on carbon nanotubes \citep{Freitag2002}.      

We use the SIM data in Figure~\ref{fig2} to quantify the shift in the Fermi energy $\Delta E_{F}$ and the variation in the density of the charge carriers $\Delta n$ with the tip voltage $V_{tip}$. The average surface potential measured from the SIM data is: $V_1=193$ mV (Figure~\ref{fig2}a), and $V_2=146$ mV (Figure~\ref{fig2}c), thus giving a difference $\Delta V = 47$ mV between the two cases corresponding to $V_{tip}=-10$ and +10V, respectively. By using the measured value for $\Delta V$ and Equation (4), we obtain the corresponding shift in the Fermi level:
\begin{equation}
\Delta E_{F} \approx e\,\Delta V = 47 \,\mathrm{meV}.
\label{F5}
\end{equation}
Moreover, for rGO in the low-energy approximation the 2D carrier density is \citep{Chen2008, Jung2008, Curtin2011}: $n \;=\; \frac{k_{F}^{2}}{\pi} \;=\; \frac{1}{\pi \,\hbar^{2} v_{F}^{2}} \bigl(E_{F}^{2}\bigr)$,  and one can calculate $\Delta n$ from the change in $E_{F}$:

\begin{equation}
\Delta n \;=\; \frac{2}{\hbar\,v_{F}} \,\sqrt{\frac{n}{\pi}}\;\Delta E_{F},
\label{F6}
\end{equation}
where $v_{F}$ is the Fermi velocity and $n$ is the density of charge carriers. For completeness, we note that the DOS--based estimate in Eq.~(6) is consistent with the capacitance analysis in Materials and Methods: taking $\Delta E_F \approx 47$~meV and $C_{\mathrm{eff}}\!\approx C_{\mathrm{geo}}$ gives a small lever arm $\eta \ll 1$, i.e.\ using $e\,\Delta V \simeq \Delta E_F$ in Eq.~(5) is appropriate in our regime where $C_Q\gg C_{\mathrm{geo}}$ \cite{Fang2007}.

We apply Equation~(6) to rGO by noting that, typically for this material, $
v_{F}^{(\mathrm{rGO})} \;\approx\; \tfrac{1}{3}\,v_{F}^{(\mathrm{graphene})}\approx\; 3.7 \cdot 10^{7}\,\mathrm{cm/s}$,
and the carrier density is very close to that of graphene $n \;\approx\; 10^{13}\,\mathrm{cm}^{-2}$ \citep{Chen2008, Jung2008, Curtin2011}. With these values and the measured $\Delta E_{F} = 47\,\mathrm{meV}$, we obtain the change in carrier density due to the tip gating:

\begin{equation}
\Delta n \;\approx\; 0.72 \times 10^{13}\,\mathrm{cm}^{-2}.
\label{F7}
\end{equation}
We emphasize that this result was obtained directly from the SIM data and the general model describing the variation of carrier with the Fermi level (Equation~(5)). Notably, a negative tip voltage increases the carrier density, indicating a p - type transport in rGO. This observation is consistent with previous reports in the literature on RGO and graphene circuits produced  via chemical and thermal reduction of graphene \citep{Kehayias2013, Jung2008, Curtin2011}.

\textit{3.2 Scanning Gate Microscopy Measurements.} To further investigate the gating effects of the voltage-biased AFM tip, we perform SGM measurements on the rGO samples. In SGM, a conductive tip with an applied voltage operates as a spatially localized gate, scanned under AFM feedback near the surface of the electrically biased rGO sample.  

\begin{figure}
    \includegraphics[width=1\linewidth]{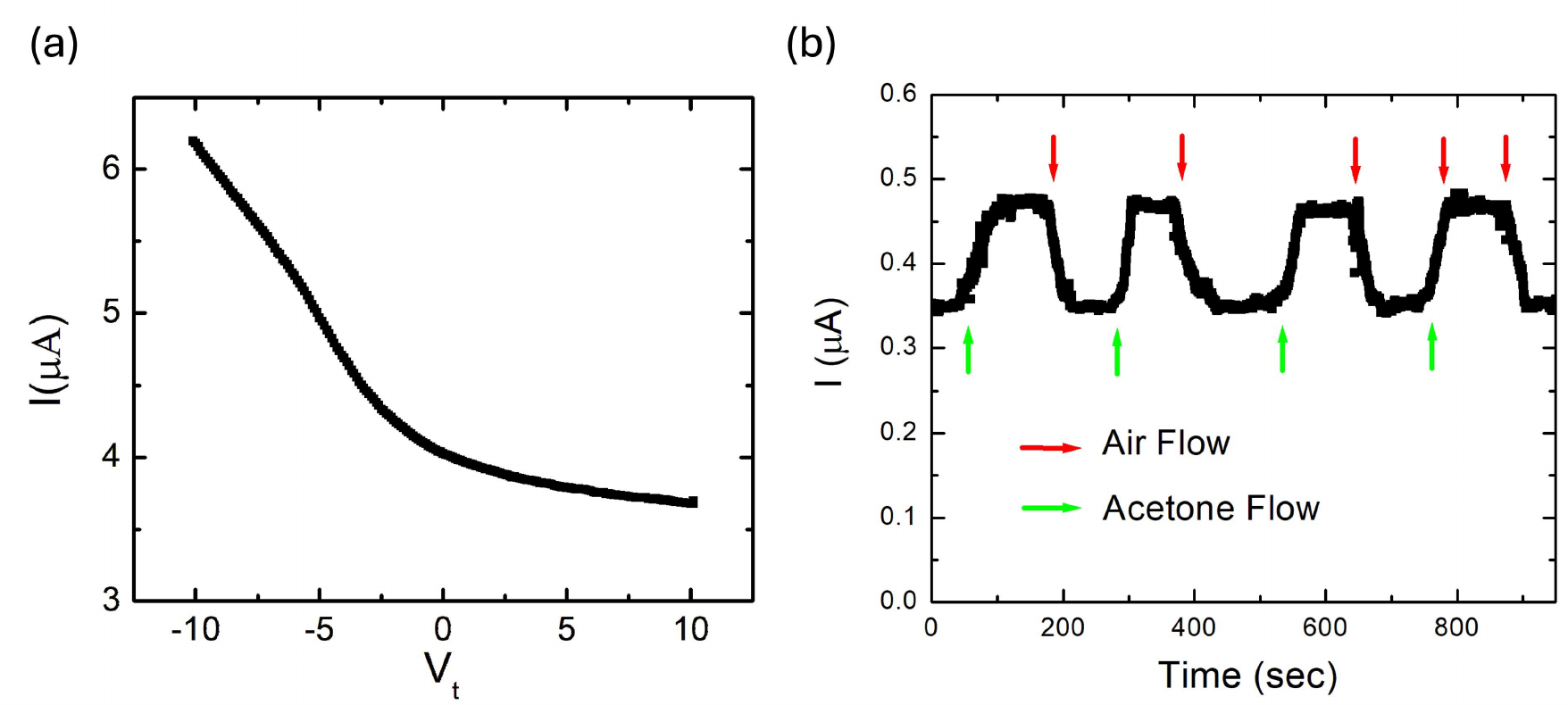}
    \caption{(\textbf{a}) Scanning Gate Microscopy data showing source-drain transport current through rGO versus the AFM tip voltage. The current through the sample decreases with increasing the tip gate voltage, consistent with p-type transport. (\textbf{b}) Chemical gating response of the same device: time trace of transport current under alternating air and analyte exposure. Upon exposure to acetone vapor (electron‑withdrawing), the current increases from 0.35~$\mu$A to 0.5~$\mu$A and returns to baseline during air purge. The cycle is repeated five times over ~15 minutes with reproducible amplitudes and recovery. \label{fig3}}
\end{figure}

Figure~\ref{fig3}a shows the transport current as a function of the tip voltage, demonstrating that rGO exhibits a strong response to the tip gate. The observed decrease in transport current from 6.2~$\mu$A to 3.7~$\mu$A as the tip voltage increases from $-10$ V to $+10$ V further confirms p - type transport in these samples. From Figure~\ref{fig3}a, we determine a ratio of approximately $1.7$ between the maximum and minimum current through the rGO sample. 

\textit{3.3 Chemical Sensing with rGO.} When the device is exposed to acetone vapor (an electron-withdrawing molecule), the source-drain current increases reproducibly from \(\sim 0.35~\mu\mathrm{A}\) to \(\sim 0.5~\mu\mathrm{A}\) and returns to baseline upon purging with air. This air-acetone cycle, repeated five times over about 15 minutes, shows consistent amplitudes and recovery (Figure~\ref{fig3}b). The sign of the response, an increase in current for an electron acceptor in a p-type channel, is characteristic of chemical gating in rGO \citep{Kehayias2013}.

Under chemical gating at fixed bias and geometry, the channel conductivity follows \(\sigma = e\,\mu\,n\). For small changes where the mobility \(\mu\) and contact resistance are approximately constant, the relative change in carrier density tracks the relative change in current $\Delta n/n \;\approx\; \Delta I/I_{air}$, where $I_{air}$ is the baseline current in air. Using the amplitudes in Figure~\ref{fig3}b, we obtain $\Delta n/n  \;\approx\;  0.43$, that is, a ~ $40-45\%$ increase in carrier density during exposure to acetone. This magnitude and sign are consistent with  electron-withdrawing dopants acting on a p-type rGO channel \citep{Kehayias2013, Deng2012, Dua2010, Robinson2008}.

The p--type response and chemical sensitivity inferred here align with our Scanning Impedance and Scanning Gate local measurements: SIM shows that tip-induced surface-potential shifts map directly to Fermi--level shifts (Equation~(5)), while SGM demonstrates that positive local gate potentials deplete holes and reduce current (Figure~\ref{fig3}a) . Together, these observations support a picture in which acetone adsorption increases the hole density in rGO, thereby enhancing conductance during exposure and recovering upon air purging.

\textit{3.4 Theoretical model of resonant scattering.} To interpret our findings, we adopt a resonant scattering model in which 
defects and impurities in the rGO lattice act as scattering centers 
when the Fermi energy lies near localized impurity states \citep{Datta1995}. In our experiments, by selectively 
shifting the Fermi level using a voltage-biased tip, we traverse regions 
of higher or lower scattering, represented in the measured peak and valley 
currents in Figure~\ref{fig3}.

\begin{figure}
\centering
\subfloat[\centering]{\includegraphics[width=0.5\linewidth]{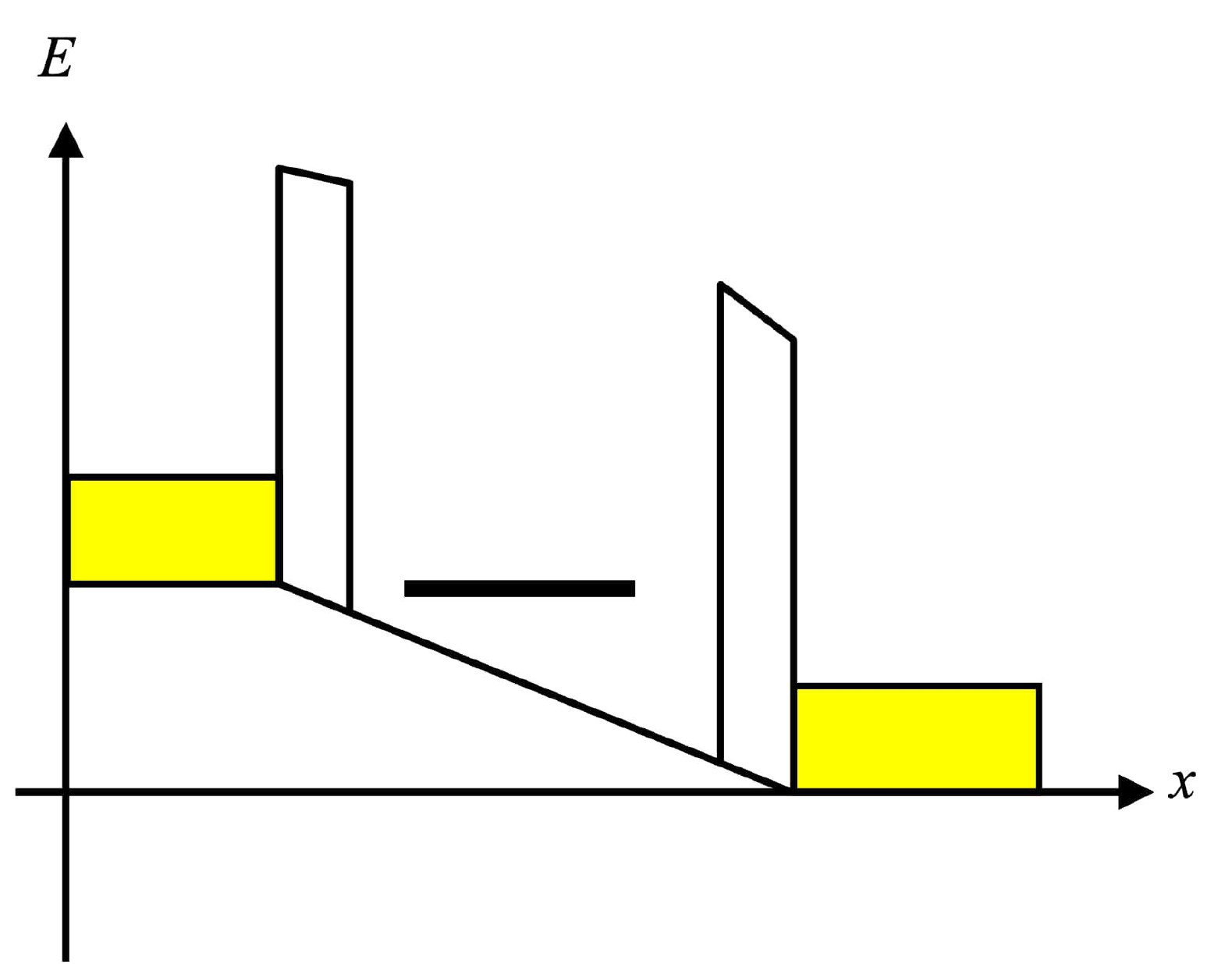}}
\subfloat[\centering]{\includegraphics[width=0.5\linewidth]{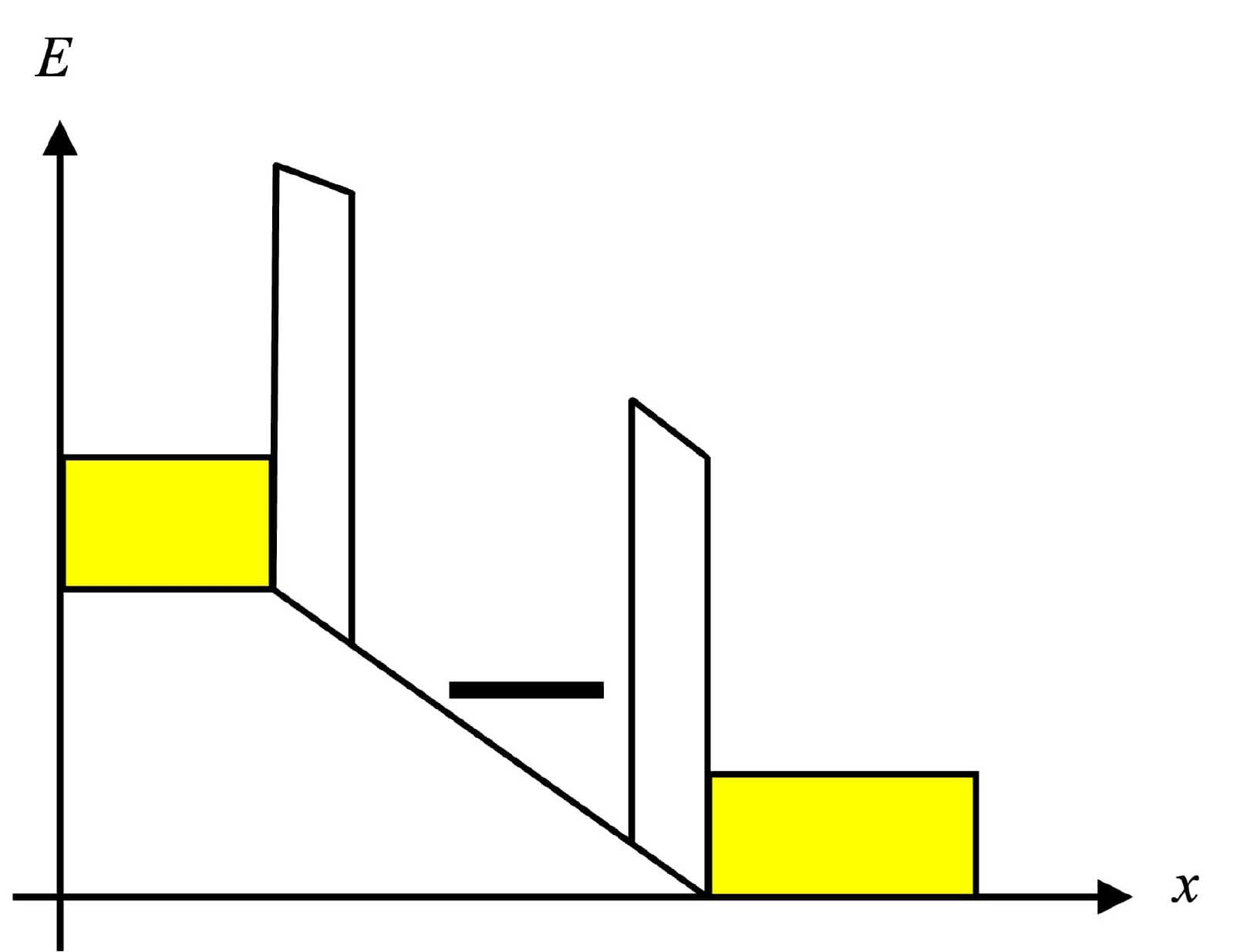}}
\caption{Schematic of energy diagrams for resonant (\textbf{a}), and off resonant (\textbf{b}) scattering of charge carriers in a double - barrier transport model. A resonant scattering state is represented by the horizontal solid bar. The available energy range for charge carriers is $\mu_{2} < E < \mu_{1}$. The lower and upper limits of this range are represented schematically by the yellow boxes. The model is discussed in detail in the text. \label{fig4}}
\end{figure} 

The theoretical model consists of two barriers (or scatterers) in series (Figure~\ref{fig4}). We can write down the transmission probability $T$ as a function of energy $E$ for the double-barrier structure as \citep{Datta1995}:
\begin{equation}
T(E) \;=\;
\frac{T_{1}\,T_{2}}{\bigl[\,1 - \sqrt{R_{1}\,R_{2}}\bigr]^{2} \;+\; 2\,\sqrt{R_{1}\,R_{2}}\;\bigl(1 - \cos\theta\bigl(E\bigr)\bigr)},
\label{F8}
\end{equation}
where $\theta$ is the the phase shift acquired in a round - trip between the scatterers, $T_{1}$ and $T_{2}$ are the transmission probabilities through barriers 1 and 2 respectively, and  $R_{1} \approx 1 - T_{1}$, $R_{2} \approx 1 - T_{2}$ are the corresponding reflection probabilities. Close to a sharp resonant level $E_r$ this expression can be written in the form:

\begin{equation}
T(E) \;\approx\;
\frac{\Gamma_{1}\,\Gamma_{2}}{\bigl(E - E_{r}\bigr)^{2} + \Bigl(\frac{\Gamma_{1} + \Gamma_{2}}{2}\Bigr)^{2}},
\label{F9}
\end{equation}
where $\Gamma_{1} = \frac{dE}{d\theta}\,T_{1}$ and $\Gamma_{2} = \frac{dE}{d\theta}\,T_{2}$. Physically, $\Gamma_{1}$ and $\Gamma_{2}$ (each divided by $\hbar$) represent the rate at which a carrier is transmitted through the double - barrier \citep{Datta1995}.

The total transmission current is obtained by integrating the transmission function over the available energy range: $\mu_{2} < E < \mu_{1}$ as indicated in Figure~\ref{fig4}. The maximum (peak) current is then given by \citep{Datta1995}:

\begin{equation}
I_{P} \;=\;
\frac{2\,e}{h} \,\int_{\mu_{2}}^{\mu_{1}}\!\!{T}(E)\,dE
\;=\;
\frac{2\,e}{\hbar}\,\frac{\Gamma_{1}\,\Gamma_{2}}{\Gamma_{1} + \Gamma_{2}}.
\label{F10}
\end{equation}

When the local AFM tip shifts $E_F$ toward a defect resonance (impurity or a structural defect in rGO), the mode transmission probability is reduced by enhanced backscattering, which lowers the net current even if the total number of transverse modes is unchanged. For our p-type device, negative $V_{\rm tip}$ increases hole density and simultaneously detunes $E_F$ from the resonance (higher $I$), while positive $V_{\rm tip}$ depletes holes and aligns $E_F$ closer to the localized level (lower $I$). As a result, the current through the sample remains close to $I_{P}$ when it is off-resonant but decreases to a "valley" current $I_{V}$ when it is resonant (see Figure~\ref{fig4}). The total valley current can be found following the same analysis as for the non - resonant case. Assuming  the linewidth is much smaller than the energy difference $E-E_r$ between the incident charge carriers and the resonant energy, we get \citep{Datta1995}: 

\begin{equation}
I_{\mathrm{V}}
= \frac{2\,e\,g\,\Gamma_{1}\,\Gamma_{2}}{\hbar \,\bigl(g\,\Gamma_{1} + \Gamma_{2}\bigr)},
\label{F11}
\end{equation}
where $g$ is a dimensionless parameter that measures the relative strength of resonant to nonresonant paths in our double–scatterer picture \citep{Datta1995}. Equations (10) and (11) imply that when scattering is highly asymmetric, with $\Gamma_{2} \ll \Gamma_{1}$, the peak and valley currents should be nearly equal. However, this is not observed in our experiments. Therefore, we assume a symmetric double - barrier $\Gamma_{1} \approx \Gamma_{2} = \Gamma$ and calculate $g$ from the experimental data in Figure~\ref{fig3} and Equations (10)-(11):  

\begin{equation}
\frac{I_{\mathrm{P}}}{I_{\mathrm{V}}}
\;\approx\;
\frac{g+1}{g}
\;=\;
1.7 \implies g \approx 1.4 
\label{F12}
\end{equation}
A larger $g$ means that resonant modes contribute more strongly to overall transport, decreasing the ratio between the peak and the valley currents. the corresponding on-resonance broadening is: 
\begin{equation}
I_{P} \;=\;
\frac{2\,e}{\hbar}\,\frac{\Gamma}{2} \implies \frac{\Gamma}{\hbar} \approx 4 \cdot 10^{13}   \, \, \mathrm{s^{-1}}
\label{F13}
\end{equation}
which provides an order-of-magnitude estimate of the defect--induced level width under our conditions.

\section{IV. Discussion}

We conducted combined scanning probe microscopy and transport measurements on reduced graphene oxide devices. Scanning gate microscopy (SGM) measurements using local gating by an AFM tip demonstrate a distinct p-type behavior in these samples. This is demonstrated by the larger current (6.2~$\mu$A) observed 
when the tip is biased at $-10$~V, compared to the smaller current (3.7~$\mu$A) at 
$+10$~V. Scanning Impedance Microscopy (SIM) experiments enable the measurement of the local surface potential and indicate a shift in the Fermi energy of 
approximately 47~meV between these two electrostatic gating conditions. This energy shift corresponds to 
a carrier density change of about $0.72 \times 10^{13}$~cm$^{-2}$. The observed variation in carrier density highlights the sensitivity of rGO's transport properties 
to local electrostatic potentials and emphasizes the crucial role of scattering 
mechanisms in determining low-dimensional transport behavior in this material. The observed p-type doping is consistent with earlier reports on chemically or thermally 
reduced graphene oxide \citep{Sehrawat2018, Kehayias2013, Jung2008, Curtin2011} and further suggests that the defects introduced 
during reduction play an active role in determining the transport properties.

To interpret these results, we consider a resonant scattering model that occurs when specific electronic states within the rGO lattice are nearly degenerate with the Fermi level. Under these conditions, impurities or structural defects in the rGO can act as resonant scatterers. Shifting the Fermi energy closer to these resonant states through local tip gating enhances scattering processes, thereby reducing the net current through the sample. Conversely, moving the Fermi energy away from these resonances decreases scattering, resulting in a higher current. The variation in transport current with the applied tip voltage, as measured by SGM, thus reflects the system's transition into and out of resonance conditions. In our experiments, applying $-10$~V to the AFM tip shifts the Fermi energy to a position where resonant scattering is less significant, leading to the observed peak current. In contrast, applying $+10$~V brings the Fermi level closer to a regime with more pronounced resonant scattering, thereby reducing the current to a lower (valley) value.

This interpretation is consistent with previous studies of doping-induced 
variations in the transport properties of graphene and rGO. These studies show that shifting the Fermi level through chemical modifications can either enhance or suppress scattering, depending on the specific configurations of impurities within the sample \citep{Geim2007, Chen2008, Wang2010, Urade2023, Jung2008, Curtin2011, Jalilian2011}. Strong evidence for resonant scattering has also been reported in studies on carbon nanotubes. For example, reference \citep{Bockrath2001} investigated one-dimensional scattering in carbon nanotubes, emphasizing electron-electron interactions within the framework of a Luttinger liquid. The study demonstrated that local gating can align discrete electronic states in the nanotube with the Fermi level, leading to either resonance-enhanced conductance or its suppression. Additionally, phonons introduce inelastic scattering processes that influence both the shape and position of resonance peaks in conductance. In particular, inelastic scattering allows charge carriers to traverse impurity sites at energies different from their initial energy, thereby expanding the range of available conduction channels.  

Our work demonstrates that, despite its two-dimensional nature, rGO can also exhibit resonant effects driven by localized impurity or defect states. rGO is a disordered two--dimensional Dirac material. In this class of systems, local gating primarily tunes the alignment of $E_F$ relative to defect/impurity resonances and charge islands, thereby modulating the transmission through inhomogeneous regions rather than invoking 1D Luttinger--liquid behavior. Inelastic electron--phonon processes broaden impurity resonances and can shift their apparent energy. Consequently, the depth and position of the current minima depend on both the local defect spectrum and phonon coupling strength. Our results suggest that the interplay between tip-induced local shifts in electrostatic potential and phonon-mediated scattering processes likely determines the precise magnitudes of the current peak and valley. In the p-type regime of our device, applying $-10$~V to the tip drives the Fermi level where resonant scattering is partially suppressed, resulting in a higher current. Conversely, applying $+10$~V shifts the Fermi level closer to resonant conditions, leading to a reduction in the measured current. Whether these resonances arise solely from localized defects, charged impurities, or a combination of defect-related and phonon-assisted processes remains an open question. Future research will focus on identifying the exact nature of these scattering centers by using spatially resolved spectroscopies or low-temperature measurements to disentangle the contributions of various scattering channels. Additionally, temperature-dependent measurements and phonon engineering approaches, such as selecting substrates that modify phonon spectra, could provide further insights into the balance between elastic and inelastic scattering processes. 

\section{V. Conclusions}

In this work, we have demonstrated that the transport properties of reduced graphene oxide can be substantially modified by the electrostatic potential induced by an atomic force microscope conducting tip. These results highlight 
the critical interplay between the density of charge carriers, defect-induced resonant states, 
and sample conductivity. In the same devices we also observe reversible chemical gating and sensing: exposure to acetone (an electron withdrawing molecule) increases the source to drain current and the current returns to baseline in air, consistent with an increase in carrier density within a simple transport picture. Moreover, the high degree of tunability 
achieved through local gating underscores the potential of rGO-based 
devices for applications requiring adaptive control of transport 
properties. Overall, the ability to modulate the conduction characteristics of rGO 
through tip-induced gating, combined with its pronounced sensitivity to 
scattering by local impurities, demonstrates the versatile nature of rGO's 
electronic structure. Further investigations into the detailed interaction between 
resonant scatterers, phonon coupling, and carrier density could provide deeper insights 
into optimizing rGO-based electronic devices.

\begin{acknowledgments}
We thank C. Rizzo and P.J. Moore for their help with sample preparation. We also thank Prof. S. Sonkusale for his assistance during the initial stages of this project.
\end{acknowledgments}

\bibliography{rGOArchives_references}

\end{document}